# Spin Dynamics Simulation of the Magneto-Electric Effect in a Composite Multiferroic Chain

Zidong Wang (王子东) and Malcolm J. Grimson

*Department of Physics, University of Auckland, Auckland 1010, New Zealand.*

A composite multiferroic chain with an interfacial linear magneto-electric coupling is used to study the magnetic and electric responses to an external magnetic or electric field. The simulation uses continuous spin dynamics through the Landau-Lifshitz-Gilbert equations of the magnetic spin and the electric pseudo-spin. The results demonstrate an accurate description of the distribution of the magnetisation and polarisation are induced by applied electric and magnetic field, respectively.

## 1. Introduction

Multiferroic materials, i.e., materials exhibit more than one ferroic (magnetic, electric, or elastic) state [1], particularly the ferroelectric (FE) and the ferromagnetic (FM) composited order is currently received intensive investigation [1-3]. In this present work, we developed the theoretical spin dynamic simulation of a one-dimensional multiferroic composite chain, which coupled by FM and FE orders. The external field driven dynamics of the magnetisation and the electric polarisation of a FM/FE system that shows a magneto-electric (ME) coupling at the interface [2,3]. The ME coupling can induce the ME effect, which is the phenomenon of inducing electric polarisation (magnetisation) by applying a magnetic (electric) field. The ME effect in composite multiferroic materials results by the combination of magnetostrictive and piezoelectric effects [4]. This can be written in a simple form,

$$ME = \frac{electric}{mechanical} \times \frac{mechanical}{magnetic} \quad or \quad ME = \frac{magnetic}{mechanical} \times \frac{mechanical}{electric} \tag{1}$$

For this purpose we considered a two-component multiferroic chain consisting of $N_S$ localized magnetic moments and $N_P$ polarisation sites. The schematic view is in Fig. 1.

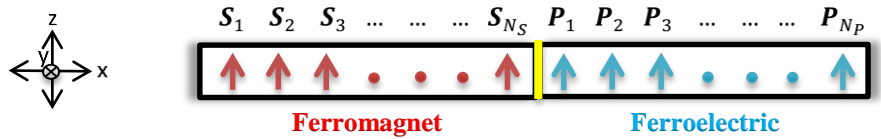

Fig. 1. (colour online) A schematic of the composite multiferroic sample built of FM and FE chains. The red and blue arrows indicate the magnetic spins and the electric pseudo-spins [5,6], respectively. The interface between FM/FE chains is indicated by a yellow line.

The FM part of the chain is a normal metal (e.g. iron, cobalt or nickel), whereas the FE part is, for example, BaTiO₃ or PbTiO₃ [3]. The total energy of the composite multiferroic system in general one-dimensional consists of three parts,

$$H = H_S + H_P + H_{SP} \tag{2}$$

where $H_S$ is the conventional Heisenberg Hamiltonian describes the FM part of the multiferroic chain with $N_S$ magnetic spins,

$$H_S = \sum_{<i,j>}^{N_S} \left( -J_S \overrightarrow{S_i} \cdot \overrightarrow{S_j} - K_S (S_i^z)^2 \right) - B(t) \sum_{i=1}^{N_S} S_i^z \tag{3}$$

where $J_S$ is the nearest neighbour exchange coupling and $K_S$ is the $z$-directional uniaxial anisotropy constant. The magnetisation vector $\overrightarrow{S_i} = (S_i^x, S_i^y, S_i^z)$ at site $i = 1, \ldots, N_S$, with the normalization $|\vec{S_i}| = 1$, and $\boldsymbol{S_j} = \boldsymbol{S_{i-1}} + \boldsymbol{S_{i+1}}$ denotes the sum of neighbour spins. The last term in (3) shows the Zeeman energy induced by the magnetic spins and an external magnetic field, $B(t)$, this field is applied alone the $z$-axis and has the time dependent form. The classical Heisenberg Hamiltonian describes the FE part of the multiferroic system with $N_P$ pseudo-spins that represents the interacting dipoles,

$$H_P = \sum_{<k,j>}^{N_P}(-J_P \overrightarrow{P_k} \cdot \overrightarrow{P_j} - K_P(P_k^z)^2) - E(t)\sum_{k=1}^{N_P} P_k^z \tag{4}$$

where $\overrightarrow{P_k} = (P_k^x, P_k^y, P_k^z)$ is a component of a pseudo-spin vector at site $k = 1, \ldots, N_P$, denotes as the directional electric dipole moment [5,6], the amplitude of the pseudo-spin vector is set to be $|\overrightarrow{P_k}| = 1$. $J_P$ is the nearest exchange interaction coupling and $K_P$ is the $z$-directional uniaxial anisotropy constant in the FE part. The system is subject to an external electric driving field, $E(t)$, that couples to the pseudo-spins in the system. The interface effects between the magnetic spin and the electric dipole systems are described by the dipole-spin interaction Hamiltonian (5), with a linear ME coupling, $g$ [3]. $\tag{5}$

$$H_{SP} = -g(\boldsymbol{S_{N_S}} \cdot \boldsymbol{P_1})$$

## 2. Spin Dynamics Method

To describe the magnetisation dynamics in the FM, a dynamic equation of spins named Landau-Lifshitz-Gilbert (LLG) equation, has been used at the atomic level [2],

$$\frac{\partial \boldsymbol{S}}{\partial t} = -\gamma'_{FM}[\boldsymbol{S} \times \boldsymbol{H}_S^{eff}(t)] - \lambda_{FM}\left[\boldsymbol{S} \times [\boldsymbol{S} \times \boldsymbol{H}_S^{eff}(t)]\right] \tag{6}$$

where $\gamma'_{FM} = \frac{\gamma}{1+\alpha_{FM}^2}$, $\gamma$ is the gyromagnetic ratio and $\alpha_{FM}$ is the dimensionless damping factor. $\lambda_{FM} = \frac{\gamma \alpha_{FM}}{1+\alpha_{FM}^2}$ denotes Gilbert damping term. $\boldsymbol{H}_{S_i}^{eff} = \frac{\partial H_S}{\partial S_i}$ is the effective magnetic field, which is a derivative of the system Hamiltonian with respect to the magnetisation, acting on each magnetic spin.

In the FE part, we used a simple pseudo-spin model [5,6] to describe the locations of the electric dipole. Since an electric dipole is a separation of positive and negative charges, a measure of this separation gives the magnitude of the electric dipole moment, it is a scalar. In the spin dynamics system, no precession of the pseudo-spins is expected (i.e., $\gamma'_{FE} = 0$) and the polarisation dynamics are described [7],

$$\frac{\partial \boldsymbol{P}}{\partial t} = -\lambda_{FE}\left[\boldsymbol{P} \times [\boldsymbol{P} \times \boldsymbol{H}_P^{eff}(t)]\right] \tag{7}$$

where the $\lambda_{FE}$ is the intrinsic damping parameter for the electric pseudo-spins and the local effective electric field in each atomic plane is defined as $\boldsymbol{H}_{P_k}^{eff} = \frac{\partial H_P}{\partial P_k}$.

## 3. Numerical Results

To demonstrate the response of the FM/FE chain to a driving field, we used the composite multiferroic model as shown in Fig. 1, with 50 spins on each side (e.g., $N_S = N_P = 50$), the nearest exchange interaction coupling $J_S = J_P = 100$, the $z$-directional uniaxial anisotropic constant $K_S = K_P = 0.01$, the ME coupling $g = 1$, the normalised gyromagnetic ratio $\gamma'_{FM} = 1$, and the damping factor $\lambda_{FM} = \lambda_{FE} = 0.1$; the applied driving field is dynamic sinusoidal type, either magnetic $B(t)$ or electric $E(t)$ field, with a magnitude of 10.

The numerical results obtained by a fourth order Runge-Kutta method and the magnetic and the electric responses were presented in Fig. 2. Top two panels show the FM/FE chain driven by a magnetic field; the mean magnetisations in the $x$-, $y$- and $z$-components are gained in Fig. 2(a), and Fig. 2(b) shows the mean magnitude of polarisation in each component. By using the same method, Fig. 2(c) and (d) show the magnetic and the electric responses under an electric driving field. In general, the electric driving field gives a quicker response than the magnetic driving field.

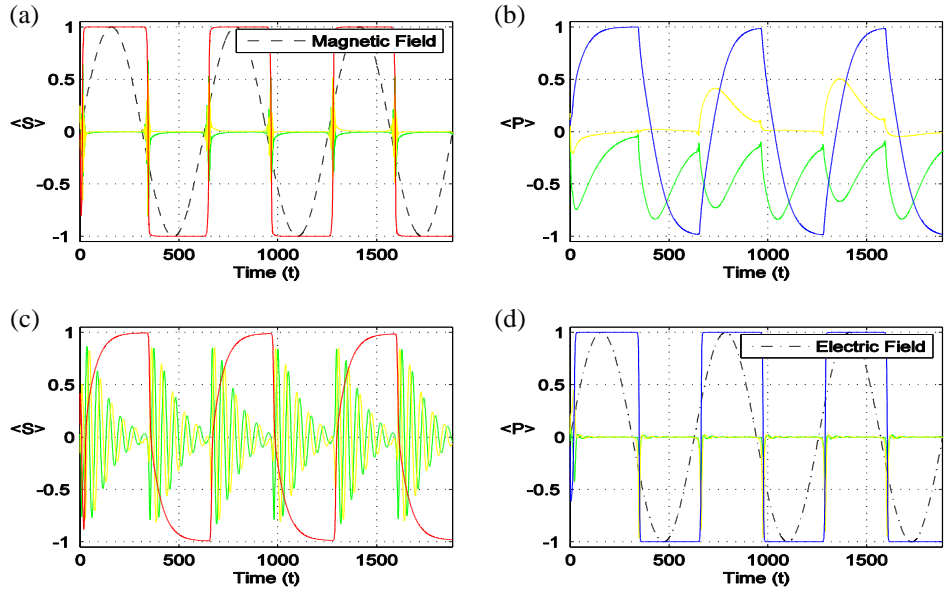

Fig. 2. (colour online) Dynamic magnetic and electric responses in the FM and FE chains, respectively. Panels (a) and (b) show the mean magnetisation (red) in the FM and the mean polarisation (blue) in the FE to an external magnetic field (black dash); (c) and (d) show the similar results as (a) and (b), but driven under an electric field(black dot dash). Green and yellow curves represent the *x*- and *y*-components, respectively.

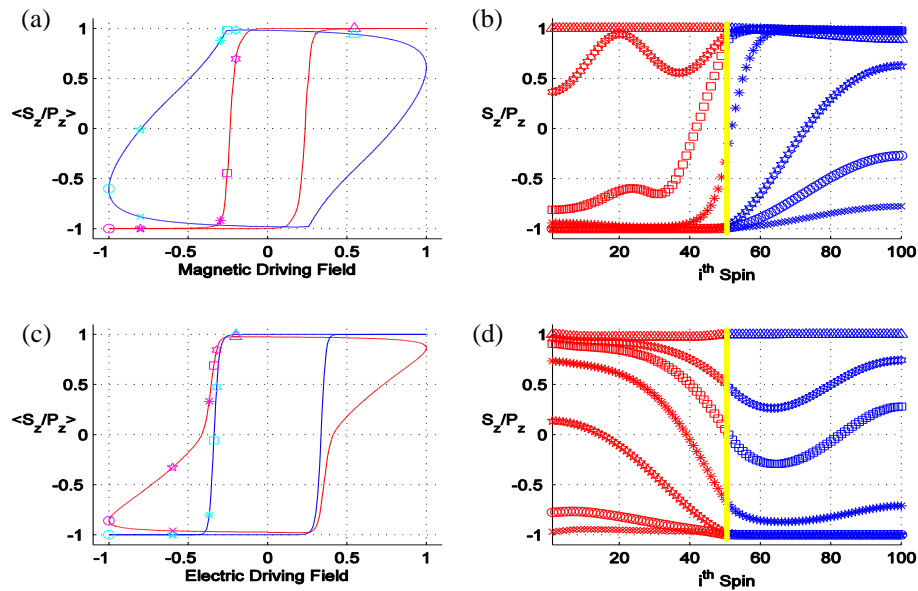

Fig. 3. (colour online) The *z*-component hysteresis loops and the spin waves at seven specific moments. The mean magnetisation and polarisation are indicated by red and blue, respectively. Each type of symbol represents a different time. The yellow line in each panel (c) and (d) shows the interface between FM/FE chains.

In order to verify the behaviour of response, a closer inspection of the FM/FE chain at seven specific moments is shown in Fig. 3. In Fig. 3(a) and (c), the magnetic (red) and electric (blue) hysteresis loops present the *z*-component mean responses contained in Fig. 2. The magnetic spins in the FM part are directly driven by the applied magnetic field is presented in left hand side of Fig. 3(b); on the other side, the electric pseudo-spins catch up slowly, given that they are only driven by the interfacial spins via the ME effect. A similar effect is displayed in the system with an electric driving field in Fig. 3(d). In Fig. 3(b) and (d) further show that magnetic spins with full precession have greater flexibility than the electric pseudo-spins without precession. Free boundary conditions at both end of the composite chain are applied in these simulations.

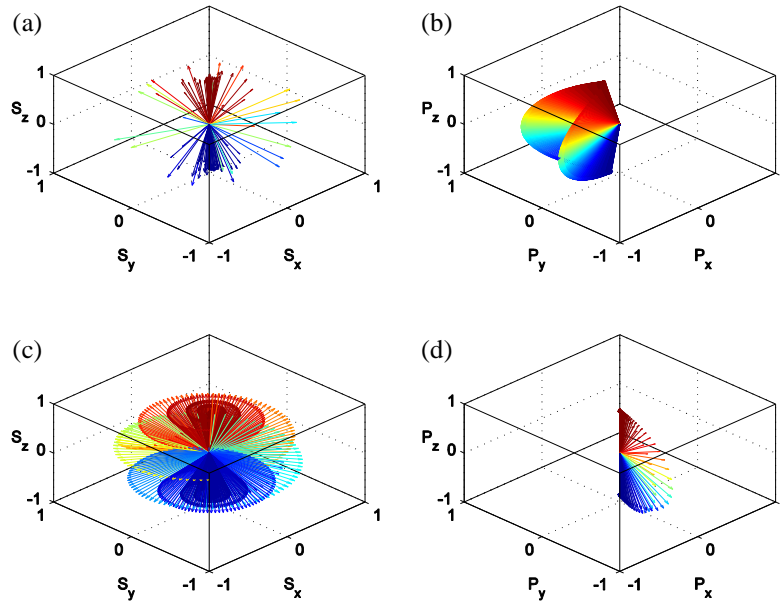

Fig. 4. (colour online) Trajectory of the particular spins in a cycle of the magnetic (top panels)/electric (bottom panels) driving field. The magnetic spins, (a) and (c) are from the FM part and the electric pseudo-spins, (b) and (d) are from the FE part. Multiple colours represent different magnitudes alone the $z$-axis.

The spin dynamics method allows us to follow the trajectories of the spins in the FM/FE chain. In Fig. 4(a) and (c), we follow a representative magnetic spin located in the bulk FM over a cycle of the driving field. Also, we show the similar behaviour for one electric pseudo-spin in the bulk FE in Fig. 4(b) and (d). The precession shown by the magnetic spin is remarkably different from the behaviour shown by the electric pseudo-spin.

## 4.    Conclusion

In this paper, the ME effect has been demonstrated by the spin dynamic method in a 1-D composite multiferroic chain. This work used the classical Heisenberg model in both FM and FE sections. As proof of concept, the response of the pseudo-spins shows a kind of flipping behaviour with respect to the electric dipole moments. Additionally, the study of the composite multiferroic system can also be done by a Monte Carlo approach. The modelling results are consistent for both methods [2,8].

## Acknowledgments

The author gratefully acknowledges Zhao BingJin & Wang YuHua for financial support.

## References

[1]    Schmid H 1994 *Ferroelectrics* **162** 317.

[2]    Sukhov A, Jia C L, Horley P P and Berakdar J 2010 *J. Phys.: Condens. Matter* **22** 352201.

[3]    Chotorlishvili L,  Khomeriki R, Sukhov A, Ruffo S and Berakdar J 2013 *Phys. Rev. Lett.* **111** 117202.

[4]    Nan C W 1994 *Phys. Rev. B* **50** 6082.

[5]    de Gennes P G 1963 *Solid State Commun.* **1** 132.

[6]    Elliott R J and Young A P 1974 *Ferroelectrics* **7** 23.

[7]    Liao J W, Atxitia U, Evans R F L, Chantrell R W and Lai C H 2014 *Phys. Rev. B* **90** 174415.

[8]    Wang Z and Grimson M J 2015 *Eur. Phys. J. Appl. Phys.* **70**, 30303.

**Declaration**

This paper is published in the Australian Institute of Physics (The AIP) for 39th ANNUAL CONDENSED MATTER AND MATERIALS MEETING, Wagga Wagga, NSW, Australia.

Link: http://www.aip.org.au/info/?q=content/39th-annual-condensed-matter-and-materials-meeting